\documentclass[10pt,aps,pra,groupedaddress,twocolumn,amsmath,amssymb,showpacs]{revtex4}
\usepackage[latin1]{inputenc}
\usepackage{amsmath}
\usepackage{amsfonts}
\usepackage{amssymb}
\usepackage{graphicx}
\usepackage{dcolumn}
\usepackage{bm}

\begin{document}

\title{Quantum and Classical Chaos in Kicked Coupled Jaynes-Cummings Cavities}
\author{A.L.C. Hayward}
\author{Andrew D. Greentree}
\affiliation{School of Physics, University of Melbourne, Vic 3050}

\date{\today}
	
\begin{abstract}
We consider two Jaynes-Cummings cavities coupled periodically with a photon hopping term.
The semi-classical phase space is chaotic, with regions of stability over some ranges of the parameters.
The quantum case exhibits dynamic localization and dynamic tunneling between classically forbidden regions.
We explore the correspondence between the classical and quantum phase space and propose an implementation in a circuit QED system.

\end{abstract}

\maketitle

\section{Introduction}
The Jaynes-Cummings (JC) Hamiltonian is the canonical model for atom-light interactions, describing a single confined bosonic mode interacting with a two level system (qubit).
This is sufficient to describe a wide range of phenomena in cavity Quantum Electrodynamics (QED).
Systems of coupled JC cavities, the Jaynes-Cummings-Hubbard (JCH) systems , have been suggested for a diverse range of optical applications such as an optical analog for the Josephson junction\cite{2009NatPh...5..281G} and  Q-switching\cite{su2008high}.
Networks of JC systems have also been predicted to exhibit phase transitions\cite{hartmann_strongly_2006,greentree_quantum_2006,angelakis:031805}. 

Improvements in the realization of photonic cavities in the lab  have made possible exploration of Jaynes-Cummings systems\cite{bishop_nonlinear_2009,blockade,fink} in the strong coupling regime in a variety of platforms. 
A current implementation of interest is in circuit QED, where  a superconducting optical resonator is capacitively coupled to a Cooper-pair box. 
This  is equivalent to a  single cavity mode of the EM field coupling to a two level atom. 
The advantage of circuit QED is that coherence times and atom-field coupling much greater than that can be achieved with visible and near infra-red systems. 
This makes circuit QED a potential medium for quantum computing, and already has been used to implement an 2 qubit Shor's algorithm\cite{2009Natur.460..240D}. 

The original proposals for observing quantum phase transitions in JCH systems\cite{hartmann_strongly_2006,greentree_quantum_2006,angelakis:031805} called for large numbers of identical systems.
Constructing large arrays of cavities which are sufficiently coherent  and identical poses a significant challenge.
Exploiting long coherence times can allow some analogous effects to be studied by trading large-scale phenomena for small-scale, long time phenomena.
For example, there is an isomorphism between the periodically kicked rotor and the Anderson tight binding model\cite{PhysRevLett.49.509}.
The Anderson model predicts localization for particles in a disordered lattice, and for dimension greater than three exhibits a second order phase transition between metallic and super fluid phases.
This has been recently demonstrated in the time-domain as a kicked system with cold atoms\cite{lemari_observation_2009}.

We examine the dynamics of a pair of periodically coupled kicked JC systems using both quantum and semi-classical treatments.
For two kicked coupled JC systems the semi-classical dynamics  are  non-integrable with a complicated phase space composed of regular and chaotic regions.
The quantum case exhibits similar structure, which converges to the classical as the number of excitations in the system increases.

Periodic systems, such as delta kicked rotors and tops, are widely used to study the link between classical and quantum chaos\cite{PTPS.98.287}.
Several interesting correspondences between the two regimes have been identified such as dynamic localization with regions of stability\cite{PhysRevLett.57.2883} and Lypanov exponents with entanglement generation\cite{PhysRevE.60.1542}. 
There are many open questions about the nature of quantum systems with semi-classical dynamics that exhibit chaotic behavior, particularly in time varying systems\cite{PhysRevA.56.4045}.

\begin{figure}
  \includegraphics[width=8cm,height=3cm]{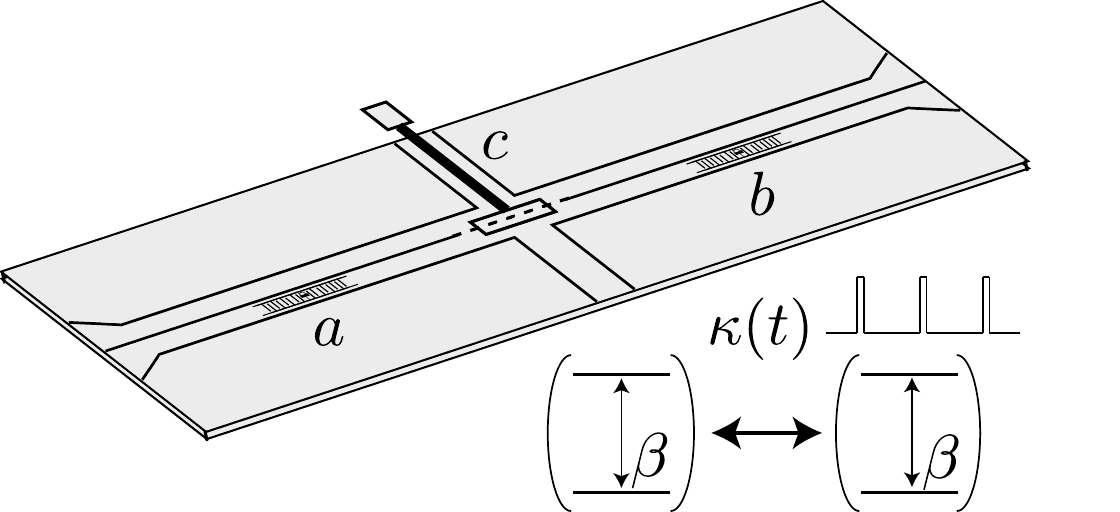}
\caption{\label{fig:cartoon}(color online) Schematic of a possible superconducting stripline cavity implementation of the kicked system. 
Transmon qubits are centered on each cavity at \textbf{a} and \textbf{b} with an atom-photon coupling $\beta$, and with the inter-cavity coupling, $\kappa$, controlled by an applied voltage at \textbf{c}. }
\end{figure}

We discuss a possible experimental implementation (figure \ref{fig:cartoon}) in a circuit QED system, compatible with the current state of the art and thus allowing an experimental investigation of quantum chaos effects in a fast developing field.
Superconducting strip-line  cavities coupled to transmons provide a JC couplingwell into the strong coupling regime\cite{bishop_nonlinear_2009}, and the architecture provides a simple means for producing the kicked coupling ($\kappa$) through an intermediate qubit\cite{PhysRevB.73.094506}.

\section{Model}
The JC Hamiltonian, in the rotating wave approximation is 
\begin{equation}
  H^{JC} = \Delta\sigma{}^{\dagger}\sigma + \beta{}(\sigma^{\dagger}a+\sigma{}a^{\dagger}) \mathbb{.}
\end{equation}
with $\sigma{}(a)$ the atomic(bosonic) annihilation operator, $\Delta$ the atom-photon detuning, coupling energy $\beta$ and we set $\hbar{}=1$.
$H^{JC}$ commutes with the total excitation number operator, $L=a^{\dagger}a + \sigma^{\dagger}\sigma$\cite{mel}. Therefore the total excitations in the cavity, $l$, is a good quantum number. 

In the bare basis, the eigenstates are 
\begin{equation}
\begin{array}{c}
|+,l\rangle{}=\sin{}\theta_{l}|g,l\rangle{} + \cos{}\theta_{l}|e,l-1\rangle{}\\
|-,l\rangle{}=\cos{}\theta_{l}|g,l\rangle{} - \sin{}\theta_{l}|e,l-1\rangle{}\mathrm{,}
\end{array}
\end{equation}
where 
\begin{align}
\tan{}\theta_{l}=2\beta{}\sqrt{l}/(\Delta{}+2\chi_{l})\mathrm{,} \\
 H^{JC}|\pm{}l\rangle{} = (\pm{}\chi{}(l) - \Delta{}/2)|\pm{}l\rangle{}
\end{align}
and
\[
\chi{}(l) = \sqrt{\beta{}^{2}l + \Delta^{2}/4}
\]
 is the generalized Rabi frequency.
Note the $\sqrt{l}$ dependence in interaction strength.
The an-harmonic energy spectrum is the source of much interesting behaviour:
In JC cavities it leads to photonic blockade\cite{tian,PhysRevA.65.063804}, providing an effective photon-photon non-linearity.
In the system under consideration, the incommensurate energies result in dynamic localization, as will be shown below.

The hopping term,
\begin{equation}
K=\kappa{}(a^{\dagger}_{1}a_2 +a^{\dagger}_{2}a_1)\mathbb{,}
\end{equation}
describes an interaction between the two cavity modes which allows photons to move from one to the other with hopping rate $\kappa$, for example, via evanescent coupling in photonic crystals, or, in the case of circuit QED, capacitive or inductive coupling\cite{PhysRevLett.90.127901}.
In our model the coupling is turned on periodically at times $t=nT$ for a short duration $\tau$. Here, $T$ is the period between kicks and $n$ an integer.
If $\tau$ is sufficiently short $(\tau \ll 1/\beta{})$, then the interaction can be described by a delta function ``kick'':

\begin{equation}
  H = H^{JC}_1 + H^{JC}_2 + \delta_{T}K'
\end{equation}
where $H^{JC}_i$ are the JC Hamiltonians for cavities 1 and 2, $\delta_T$ is a periodic delta function with period $T$ and $K' = K\tau{}$.
We also require that $\tau \gg 1/\omega$, so that the rotating wave approximation is valid.

The three dimensionless parameters, $\kappa{}\tau$, $T\beta{}$ and $\Delta\beta{}$, are sufficient to specify the dynamics of $H$.
For simplicity we consider only the quasi-resonant case, $\Delta \sim 0$, where the key features of the system are most easily elucidated. This makes $\sin{\theta_{l}} = \cos{\theta_{l}} = \frac{1}{\sqrt{2}} $ in equation 2.

The coupling term breaks the individual excitation conservation of each JC system, but commutes with the total $L = L_1 + L_2$, thus we can consider cases of total excitation number individually.
For a single excitation, $L = 1$, the excitation oscillates between cavities trivially, with frequency $\kappa\tau$, and so we do not dwell on this case. 
For all $L > 1$  we find rich behavior with signatures of quantum chaos. However, here we confine ourselves to $L = 2$ in the quantum case, and the semi-classical equivalent.
Although the dimension of Hilbert space is just 8, many of the features of quantum chaos are already present, and it is this case which will be most accessible experimentally.

\subsection{Semi-Classical dynamics} We derive the classical equations of motion by taking the expectation value of the Heisenberg equations of motion (see, for example, \cite{Filipowicz:86}). 
Between kicks each system evolves separately as 
\begin{equation}
  \begin{array}{l}
 \langle{}\dot{a}\rangle=\dot{E}=-i\beta{}S, \\
\langle{}\dot{\sigma{}}\rangle{}=\dot{S}=i\Delta{}S+i\beta{}ES_z, \\
\langle{}\dot{\sigma_{z}}\rangle{}=\dot{S_z}=2\beta{}i(SE^{*}-S^{*}E)\mathrm{,}
 \end{array}
\end{equation}
where $E$, the E-field, and $S$, vectors on the Bloch sphere are now classical quantities.  
For no detuning the uncoupled equations of motion are equivalent to that of a pendulum with the momentum $E$ and $S_z = \cos{\theta}$, the height of the bob.
This motion has two constants of motion,
\begin{equation}
\begin{array}{l}
N_{i} = |E_{1,2}|^2 + \frac{1}{2}(S_{z i}+1) \\
S_z^2 + 4S^{*}S = 1\mathrm{.}
\end{array}
\end{equation}
While this has an analytical solution in terms of elliptical functions, in practice it is easier to numerically integrate.

The kick is given by the map
\begin{equation}
\left(\begin{array}{c}E_1 \\ E_2\end{array}\right)_{n+1} =
\left(\begin{array}{cc} \cos{\kappa'} & \sin{\kappa'} \\
-\sin{\kappa'} & \cos{\kappa'} \end{array}\right)
\left(\begin{array}{c}E_1 \\ E_2\end{array}\right)_n \mathbb{.}
\end{equation}
  
The kicked hopping leads to non-integrable dynamics, so that the only constant of motion is now  $N_1 + N_2 = N$. 
In general this results in a chaotic phase space, however, for some values of $\kappa$ and $T$ there will be regions in which the motion is semi-regular.
These regions are described by KAM (Kolmogorov-Arnold-Moser) theory\cite{transition}.
In an unperturbed system the path in the $d$ dimensional phase space in action-angle variables lies on the surface of a $d$-torus.
If the periods in each dimension are sufficiently incommensurate then the system is confined  near a deformed torus for  small perturbations.
The system becomes increasingly chaotic as the perturbation is turned up, leading to destruction of some tori.
The phase space is then a chaotic sea with islands of stability which are topologically separate, from the chaos as well as each other.
Eventually the perturbation destroys all these regions and the dynamics are fully chaotic.

\begin{figure}
	\begin{center}
		$\begin{array}{ccc}
 			\includegraphics[height=4cm]{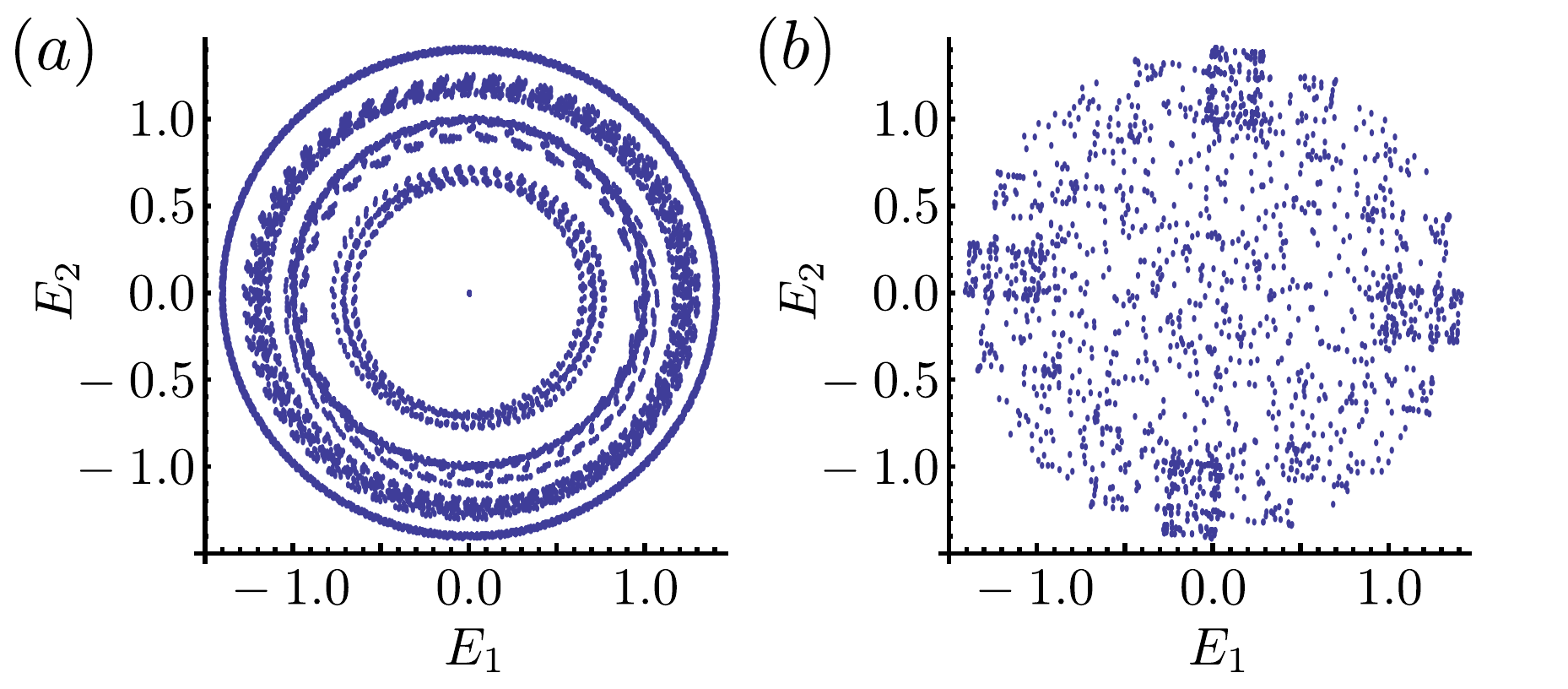}
		\end{array}$
	\end{center}

\caption{\label{fig:strobe}Classical phase space strobe plots of the E-field in each cavity over 200 kicks from several initial points.  
a) $\kappa\tau = 1.3, \beta{}T = 0.1$. 
In the small $T$ limit the total total energy in the electric field, $E_1^2 +E_2^2$ is stable, leading to non overlapping rings.
b) $\kappa\tau = 0.4, \beta{}T = 1.7$. Phase space is mostly chaotic except for the four regions where the energy in the system is confined mostly to one cavity.
As there are 4 degrees of freedom, and only a single constant of motion, plots of the electric field in each cavity do not convey the entire dynamics.
} 

\end{figure}

The centers of stability that survive the longest are usually found around short periodic orbits.
In this kicked system, however, there are in general no single-period orbits, making the motion difficult to determine the precise point at which the phase space becomes fully chaotic.
However, numerical simulations for the $N = 2$ case indicate that for small $\kappa\tau$ the most persistent KAM tori are around $N_{1,2} =\sqrt{2}\sin{(\kappa\tau)^2}, N_{2,1} = \sqrt{2}\cos{(\kappa\tau)^2}$ (Figure \ref{fig:strobe}b).
That is, in these four regions of phase space the energy in the system remains localized to a single cavity.
 Each period 
As $\kappa\tau$ is increased these regions become leaky (cantori) and eventually disappear, after which the phase space is fully chaotic.

The value of $\kappa\tau$ at which the system becomes chaotic is dependent on $T$.
 The period for a small electric field in a cavity is $2\pi{}$; when $\beta{}T$ is resonant with this the KAM tori are destroyed with much smaller $\kappa\tau$.
 Unlike  other kicked systems, this system is still regular for some $\kappa\tau$ at the resonances due to the non-linear nature of the perturbation that each cavity sees. 
The range of parameters in which this mode occurs is shown in (figure \ref{fig:param}a) where the destabilizing effect of the resonances can be seen around $\beta{}T=2n\pi{}$.

We can also consider the limit in which $\kappa\tau$ is larger then the kick period, $\beta{}T$.
 In this limit the electric field decouples from the atomic degrees of freedom and the energy in the electric field oscillates between the two cavities(figure \ref{fig:strobe}b) and we have separate regions which conserve the total energy of the field.
 For small kick period, $T \ll \beta{}$ there is a center of stability around $S_{z1}=S_{z2} = 0$, dynamically confining the atoms to their ground states.

\subsection{Quantum Dynamics} We find that the quantum dynamics exhibit some  qualitatively similar behavior to the classical case, however, there are also effects which arise which are specifically quantum in nature.

To explore these dynamics we define the Floquet operator $U_f$ which evolves the system from time $t=nT^+$ to $t=(n+1)T^+$:
\begin{equation}
U_f = e^{-i(H^{JC}_1 + H^{JC}_2)T}e^{iK} = e^{-iH_0T}e^{iK}\mathrm{.}
\end{equation}
The dynamics of a kicked system can be studied though the eigenstates, $f_i$ of $U$.
On application of $U$ the Floquet states pick up eigenphase $e^{i\lambda{}_i}$.
Thus the problem is equivalent to a time invariant Hamiltonian.
This allows the calculation of the long term behavior of the system.

The quantum equivalent of KAM tori can be understood as dynamic localization\cite{1992ASIC..357.....C}:
States which are initially in the localized regions have exponentially suppressed diffusion into chaotic areas of phase space.

If some state $\psi$ is well represented by a small number of basis states, $\psi^0_i$ we may consider $\psi$ to be localized to some degree.
 This can be quantified with the participation number($P$)\cite{meja-monasterio_entanglement_2005}:
\begin{equation}
P(\psi) = \left({}d\sum\limits_{i}^d|\langle{}\psi{}|\psi^0_i\rangle{}|^{4}\right){}^{-1}
\end{equation}
which we have normalized by the total dimension $d$ of the space. $P$  is 1/d when $|\langle{}\psi|\psi^0_i\rangle{}|=1$  for some $i$ and 1 when $\psi$ projects evenly onto the $\psi^0_i$.
 One can consider this to be a indication of quantum ergodicity\cite{1998PhyD...33...77C}.

While $P$ is dependent on the choice of basis (ie. we can always choose some basis with $\psi$ as a base), comparing the eigenstates of the unperturbed Hamiltonian to the perturbed best represents the degree of mixing\cite{PhysRevLett.71.529}.
We therefore take the $\kappa = 0$ eigenstates as the basis, and increasing $\kappa$ leads to Floquet states with increasing $P$.

\begin{figure}
	\begin{flushleft} 
		$\begin{array}{ccc}
			\includegraphics[height=4cm]{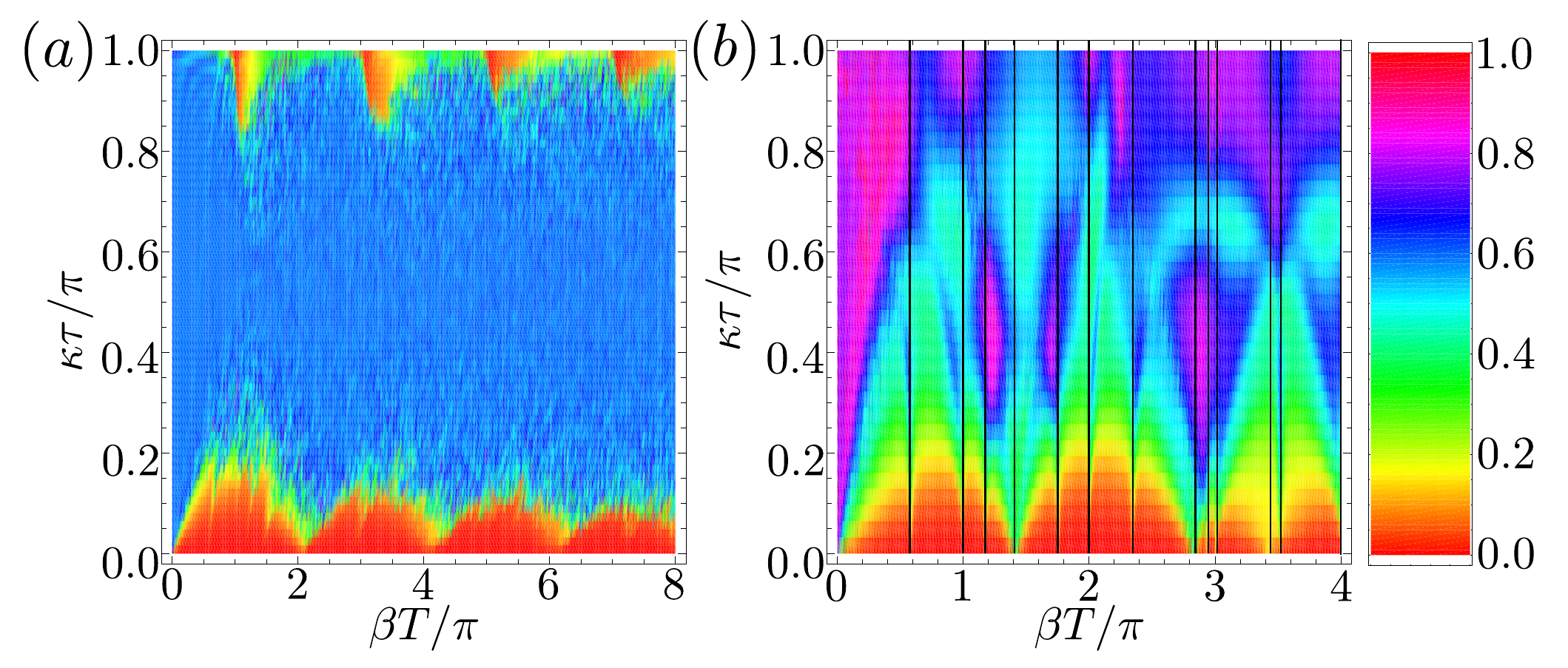}
		\end{array}$
	\end{flushleft}

\caption{\label{fig:param}(color online) a) Classical: Average $N_2$ over 1000 kicks for points initially at $E_2=\sqrt{2}$. 
The red regions represent parameters where most of the energy in the system is localized to a single cavity, ie. The chaotic border. 
The KAM tori are destroyed with relatively small kick strength at $\beta{}T = 2n\pi{}$.
b) Quantum: Average participation number of Floquet states.
The regions of localization are qualitatively similar to those in the classical case, however, note that the $T$ scale is different in each graph. Vertical lines mark the location of resonances. 
}
\end{figure}

Figure \ref{fig:param}b shows the average participation number of the Floquet states over a range of $\kappa\tau$ and $\beta{}T$ for a system with two excitations.
We denote the subspace of states with two excitations in the one cavity as $|\psi^2_i\rangle{}$s, and likewise the states with one excitation in each cavity as $\psi^1_i$s.
The regions where $P$ is small corresponds to states with both excitations in the same cavity being dynamically separated from states with excitations in both cavities, ie. an approximate symmetry of $U_f$.

The suppression is destroyed by resonances which occur at $T= \frac{t}{2\pi}=\frac{n\sqrt{2}}{2}, n(1+\frac{\sqrt{2}}{2}), n(1-\frac{\sqrt{2}}{2})$, which are solutions to
\[
\sqrt{2}T =mT,   m \in \mathbb{I} \mathrm{.}
\]
At these valuies the phase accrued after each period is 0, and so there is no destructive interference.
 This implies that is indeed dynamical localization suppressing dispersion in the system. 
For example, when  $T = \frac{n\sqrt{2}}{2}$,the states  in $|\psi^2\rangle$ pick up no relative phase to states with $E=0$. This removes the interference suppressing transmission into these states, and destroys the localization.

In figure \ref{fig:param}b) we can see, for the atomic limit, that the dependence of localization on the parameters correspond qualitatively to the semi-classical case, though with important differences.
The frequency at which the classical cavities oscillate depends continuously on the energy in the cavity, and in general is different from the Rabi frequency of the quantum case; these two only coincide in the limit $l \rightarrow \infty$.
Thus, the locations of resonances are different in the two regimes.

Note also that in contrast to the classical case, the resonance removes the localization for arbitrarily small $\kappa\tau$.
Resonances in the classical case are not sharp, due to the energy dependent frequencies.

\begin{figure}
		\includegraphics[height=4.3cm]{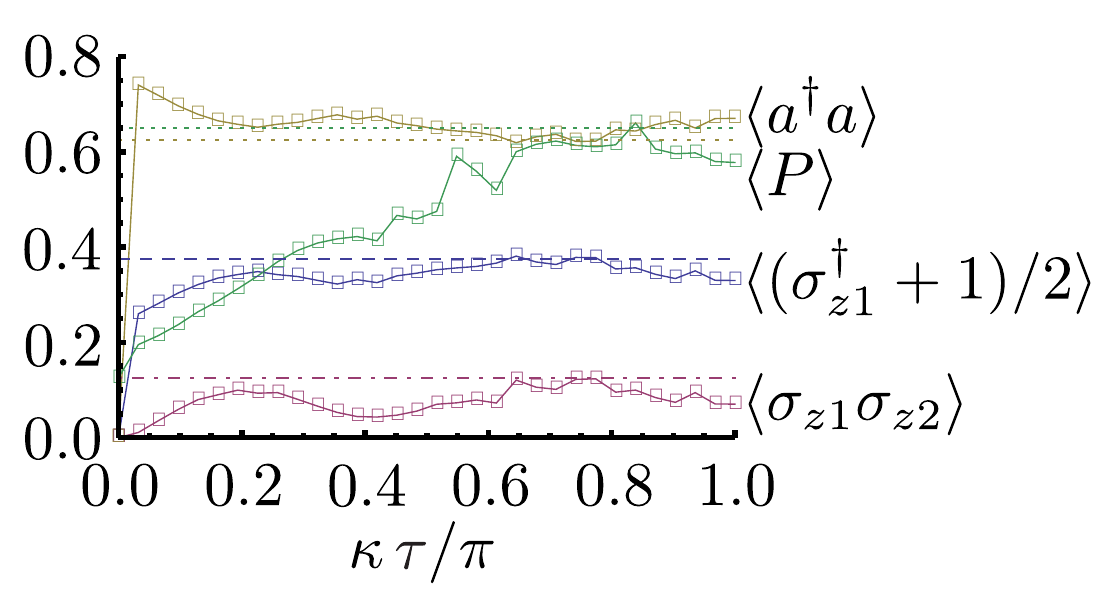}
	\caption{\label{fig:exp}Average participation number and long term averages of measurable quantities (mean expectation of an ensemble of random states) over increasing $\kappa{}\tau{}$ with $\beta{}T=1.2$. Circles(Dashed): $(\sigma^{\dagger}_{z1}+1)/2$, Diamonds(Dots): $a^{\dagger}_1a_1$, Squares(Dot-Dashed):$\sigma_{z1}\sigma_{z2}$, Triangles: Average participation number(Dot-Dashed).}
\end{figure}

For time independent systems, chaos can be studied via the statistics of energy levels, however, in periodic systems, the eigenphases of the unitary operator are not observable. Ergodicity of can be explored experientially by comparing the expectation of observables in the system to an ensemble of random states. For a chaotic system, the unitary map $U_f$ has no symmetries, and so we expect the average state to be no different from a random one chosen with the appropriate measure. Figure \ref{fig:exp} shows the long-time average of some experientially observable quantities, and the expected average of a random state. 

Classically, islands of stability are topologically separated, forbidding transitions between them.
Quantum dynamics admit such flow of probability in phase space by a mechanism called dynamic tunneling and has been observed experimentally in a variety of systems\cite{steck2}.
Although this mechanism is distinct from the usual tunneling, as there is no potential barrier to overcome, the system nevertheless moves across classically forbidden regions in phase space.

\begin{figure}
	\begin{center}
 			\includegraphics[height=3.4cm]{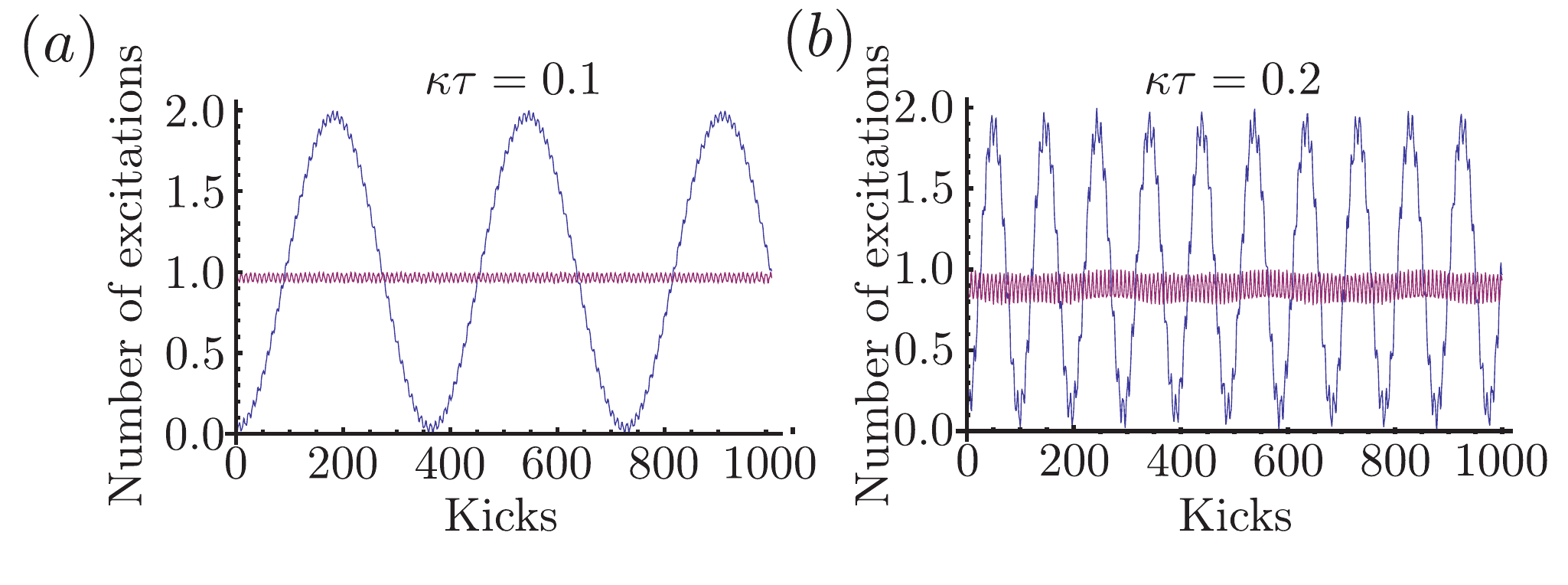} 
	\end{center}

 \caption{\label{fig:tunnel}(color online) Evolution of the system initially in $|\psi^2\rangle{}$ over 1000 kicks with $\beta{}T = 1.2$ and a) $\kappa\tau = 0.1$ b) $\kappa\tau = 0.2$. Blue line is the expectation of excitations in cavity 1. Purple line is the expectation of finding both excitations together in the one cavity.}
\end{figure}

In the $\kappa=0$ limit there is a two fold degeneracy for all Floquet states due to the $H^{JC}_1, H^{JC}_2$ symmetry.
A state, $|\psi\rangle$, initially in $|\psi^2_1\rangle$ in cavity one is in a superposition of  two Floquet states, $|\pm{}f^2\rangle{}$, which have equal projections in both cavities, but still in the $|\psi^2_i\rangle$ subspace:
\begin{equation}
|\psi\rangle = |\psi^2_1\rangle = \frac{1}{\sqrt{2}}\left({}|+f^2\rangle{} + |-f^2\rangle{}\right){}\mathrm{.}
\end{equation}
The perturbation breaks the degeneracy, leading to an approximate separation in the eigenphases, $\phi$.
Each kick, the two Floquet states composing $|\psi\rangle{}$ are separated by a phase-angle of $\phi$.
After $\frac{\pi}{2\phi}$ kicks the phase separation is $\pi$, and $|\psi\rangle$ has evolved to the state $\frac{1}{\sqrt{2}}(|+f^2\rangle{} - |-f^2\rangle{} = |\psi^2_2\rangle$, i.e. completely in the other cavity.
Figures \ref{fig:tunnel}a) and \ref{fig:tunnel}b) show the transmission between the two separated localized states for $\kappa\tau{} =  0.1$ and  $\kappa{}\tau =  0.2$ respectively. The two excitations in the system oscillate between cavities, though are strongly localized to the $\psi^2_i$ subspace. As $\kappa\tau$ increases so does $\phi$, and the localization to the $|\psi^{2}_{i}\rangle{}$ subspace decreases.

\section{Experimental Implementation}
While the effects discussed apply to any implementation of JC systems, circuit QED (cQED) presents itself as one of the most viable platforms due to the large coupling coefficients and long coherence time, relative to other cavity QED systems.

Current experiments in cavity QED, where a transmon is coupled to a resonating microwave cavity, have characteristics which could allow a successful realization of this kicked system.
A cQED setup with $\omega/2\pi = 6.92\mathrm{GHz}$, $\beta{}/2\pi = 347\mathrm{MHz}$ and coherence time  of order $1\mu{}s$ has been achieved recently\cite{bishop_nonlinear_2009,blockade}.

The localization transition occurs around $\kappa{}\tau \approx .1$ and for the delta-function kick approximation to be valid we need the pulse time $\tau \ll 1/\beta$.
For the coupling strengths cited above, this requires a pulse time of $\tau \approx 10^{-10}s$ and, therefore, $\kappa$ order $1\mathrm{GHz}$.
Between pulses $\kappa$ must be of the same order as the decoherence rate (ie. $\sim 1\mathrm{MHz}$)  such that the dispersion due to the constant inter-cavity coupling is small over the time of the experiment.
Thus a sequence of $\sim 100$ kicks could be applied within the coherence time. We have seen that this is long enough to observe dynamic tunneling and localization/delocalization by inlcuding the decoherence and dephasing explicitly in the simulation.

The tunable hopping term could be achieved using an intermediate qubit coupling such as in \cite{PhysRevLett.90.127901,PhysRevB.73.094506}. 
In such scheme's the effective coupling is of order 
\[
\kappa_{\mathrm{eff}} \sim \beta_{13}\beta_{23}/\Delta_{3}
\]
 where $\beta_{13}$, $\beta_{23}$ and $\Delta_3$  are the coupling strengths of each resonator to the intermediate qubit and it's detuning respectively and $\Delta_{3} \gg \beta{}$.
 This requires the coupling to the intermediate qubit to be significantly greater than the other couplings.
 The detuning can be controlled in situ, allowing the coupling to be switched on and off.

Spectroscopic measurements can be used to determine the final state\cite{fink}.
Although there will be significant interaction with the environment, the only final states of interest are those that still have two excitations.
One can therefore largely remove the effects of atomic relaxation and photon dissipation with a post-selection scheme, given a temperature smaller then then the characteristic energies of the system.
De-phasing terms will still be relevant, however, these are generally ignorable over the time frames considered\cite{bishop_nonlinear_2009}.

\section{Discussion/Conclusions} 
The phenomena discussed have been observed in other systems, such as dynamic tunneling and localization in cold atoms\cite{steck2,hensinger}. Circuit QED  allows direct control over many system parameters and direct measurement of the state of the system. This can be used, for example, to study the effect of noise by controlling the detuning parameter in situ.

As circuit QED is proving to be an important field, with a wide range of possible applications, understanding chaotic behavior in these systems will be crucial.
 An experimental realization of the system seems quite possible, although it is not without challenges, specifically in achieving a sufficiently large inter-cavity coupling.
It would allow the study of the rich behavior that can be expected in coupled Jaynes-Cummings systems, and open up new regimes for investigating quantum chaos.

We have presented a simple model which exhibits a transition from localization to ergodicity and dynamic tunneling. 
Importantly, we see this behavior even for small Hilbert space dimension, which,
although interesting behavior can be seen for any number of excitations above two, the lowest case most clearly conveys the aspects we have emphasized. 
Furthermore, the two excitation case will most likely be the easiest to implement experimentally.
Constantly improving control in circuit QED systems means that it will be possible to study the higher dimensional cases. 
This could potentially allow a novel means for probing the transition between classical and quantum chaos.

The authors thank T. Duty and G.J. Milburn for helpful discussions. A.D.G. acknowledges the ARC for financial Support (Project No. DP0880466)

\bibliographystyle{h-physrev3}
\bibliography{hayward}

\begin{thebibliography}{10}

\bibitem{2009NatPh...5..281G}
D.~{Gerace} {\em et~al.},
\newblock Nat Phys {\bf 5}, 281 (2009).

\bibitem{su2008high}
C.~{Su} {\em et~al.},
\newblock \pra {\bf 78}, 62336 (2008).

\bibitem{hartmann_strongly_2006}
M.~J. {Hartmann}, F.~G. S.~L. {Brandao}, and M.~B. {Plenio},
\newblock Nat Phys {\bf 2}, 849 (2006).

\bibitem{greentree_quantum_2006}
A.~D. {Greentree} {\em et~al.},
\newblock Nat Phys {\bf 2}, 856 (2006).

\bibitem{angelakis:031805}
D.~G. Angelakis, M.~F. Santos, and S.~Bose,
\newblock \pra {\bf 76}, 031805 (2007).

\bibitem{bishop_nonlinear_2009}
L.~S. {Bishop} {\em et~al.},
\newblock Nat Phys {\bf 5}, 105 (2009).

\bibitem{blockade}
K.~M. {Birnbaum} {\em et~al.},
\newblock Nature {\bf 436}, 87 (2005).

\bibitem{fink}
J.~M. {Fink} {\em et~al.},
\newblock \prl {\bf 103}, 083601 (2009).

\bibitem{2009Natur.460..240D}
L.~{Dicarlo} {\em et~al.},
\newblock \nat {\bf 460}, 240 (2009).

\bibitem{PhysRevLett.49.509}
S.~{Fishman}, D.~R. {Grempel}, and R.~E. {Prange},
\newblock \prl {\bf 49}, 509 (1982).

\bibitem{lemari_observation_2009}
G.~{Lemarie} {\em et~al.},
\newblock Arxiv, 0907.3411  (2009).

\bibitem{PTPS.98.287}
G.~{Casati} and L.~{Molinari},
\newblock Prog. Theoretical Phys. Supp. {\bf 98}, 287 (1989).

\bibitem{PhysRevLett.57.2883}
T.~{Geisel}, G.~{Radons}, and J.~{Rubner},
\newblock \prl {\bf 57}, 2883 (1986).

\bibitem{PhysRevE.60.1542}
P.~A. {Miller} and S.~{Sarkar},
\newblock PRE {\bf 60}, 1542 (1999).

\bibitem{PhysRevA.56.4045}
D.~W. {Hone}, R.~{Ketzmerick}, and W.~{Kohn},
\newblock \pra {\bf 56}, 4045 (1997).

\bibitem{PhysRevB.73.094506}
A.~O. Niskanen, Y.~Nakamura, and J.-S. Tsai,
\newblock Phys. Rev. B {\bf 73}, 094506 (2006).

\bibitem{mel}
M.~I. Makin {\em et~al.},
\newblock Phys. Rev. A {\bf 77}, 053819 (2008).

\bibitem{tian}
L.~Tian and H.~J. Carmichael,
\newblock Phys. Rev. A {\bf 46}, R6801 (1992).

\bibitem{PhysRevA.65.063804}
S.~Rebi\ifmmode~\acute{c}\else \'{c}\fi{}, A.~S. Parkins, and S.~M. Tan,
\newblock \pra {\bf 65}, 063804 (2002).

\bibitem{PhysRevLett.90.127901}
A.~Blais, A.~M. van~den Brink, and A.~M. Zagoskin,
\newblock Phys. Rev. Lett. {\bf 90}, 127901 (2003).

\bibitem{Filipowicz:86}
P.~{Filipowicz}, J.~{Javanainen}, and P.~{Meystre},
\newblock J. Opt. Soc. Am. B {\bf 3}, 906 (1986).

\bibitem{transition}
L.~Reichl,
\newblock {\em {The transition to chaos}} (Springer Berlin, 1992).

\bibitem{1992ASIC..357.....C}
P.~{Cvitanovi{\'c}}, I.~{Percival}, and A.~{Wirzba},
\newblock  {\bf 357} (1992).

\bibitem{meja-monasterio_entanglement_2005}
C.~{Mejia-Monasterio} {\em et~al.},
\newblock \pra {\bf 71} (2005).

\bibitem{1998PhyD...33...77C}
B.~V. {Chirikov}, F.~M. {Izrailev}, and D.~L. {Shepelyansky},
\newblock Physica D Nonlinear Phenomena {\bf 33}, 77 (1998).

\bibitem{PhysRevLett.71.529}
L.~{Benet}, T.~H. {Seligman}, and H.~A. {Weidenm\"uller},
\newblock \prl {\bf 71}, 529 (1993).

\bibitem{steck2}
D.~A. Steck, W.~H. Oskay, and M.~G. Raizen,
\newblock Science {\bf 293}, 274 (2001).

\bibitem{hensinger}
W.~K. {Hensinger} {\em et~al.},
\newblock \nat {\bf 412}, 52 (2001).

\end{thebibliography}

\end{document}